\documentclass[11pt]{article}

\usepackage{epsfig}
\usepackage{latexsym}
 
\newcommand{\bi}{\begin{itemize}}
\newcommand{\ei}{\end{itemize}}
\newcommand{\be}{\begin{equation}}
\newcommand{\ee}{\end{equation}}

\newcommand{\bea}{\begin{eqnarray}}
\newcommand{\eea}{\end{eqnarray}}
\newcommand{\beastar}{\begin{eqnarray*}}
\newcommand{\eeastar}{\end{eqnarray*}}

\newcommand{\eq}[1]{~(\ref{#1})}

\begin{document}
\mark{{Thermal behavior of small systems}{F. Ritort}}
\title{Single molecule experiments in biophysics: exploring the thermal behavior of nonequilibrium small systems}

\author{F. Ritort~$^{\dag}$\\
$\dag$ Departament de F\'{\i}sica Fonamental, Facultat de F\'{\i}sica,
Universitat de Barcelona,\\ Diagonal 647, 08028 Barcelona (Spain)\\
{\tt E-Mail:ritort@ffn.ub.es}}


\maketitle

\abstract{Biomolecules carry out very specialized tasks inside the cell where
energies involved are few tens of $k_BT$, small enough for thermal
fluctuations to be relevant in many biomolecular processes.
In this paper I discuss a few concepts and present some
experimental results that show how the study of fluctuation theorems applied to
biomolecules contributes to our understanding of the nonequilibrium
thermal behavior of small systems.}

\section{Biomolecules, molecular demons and statistical physics.}
Biophysics is a relatively young discipline that is becoming steadily
popular among statistical physicists~\cite{Houches02}. Although there
are several reasons behind this general upsurge of interest, a very
attractive aspect of biophysics is its strong interdisciplinary
character. In recent years biophysics is facing the dawn of an
unprecedented fusion of various knowledges coming from different
traditional scientific areas from physics to chemistry, biology and
computer science. At the root of such melting pot there is the
discovery of the molecular structure of the gene by Crick and Watson
in 1953. This has established the basis for a new ``solid state''
science in biology, a bit akin to the role played in modern solid
state and condensed matter physics by the discovery of the atom one
century ago.

The current knowledge about the cell shows it as a very complex organism
made out of several parts that carry out different specialized tasks
organized into a modular structure, a bit like a farm or factory where
different sections or departments are in charge of performing different
tasks. This modular organization is extremely complex as it consists of
different levels intertwined in a big fuss yet to be understood. The
result of all these interactions is a web of informational flow where
actions at one level trigger responses in another, cell differentiation
being a prominent example. Among these levels of complexity, molecular
biophysics is a discipline whose scope is to investigate the structure
and function of biological matter starting from the physico-chemical properties
of constituents molecules. Within this level it is nowadays
possible, thanks to the development of nanotechnologies, to experimentally 
manipulate individual molecules while they carry out specialized
molecular functions. In single-molecule experiments the information
that can be gathered is fundamentally kinetic as molecules can be
individually followed in time during a process which
often occurs out of equilibrium. The merge of this knowledge with the
static information gained from structural biology studies provides a
promising framework to elucidate the function of many biomolecules.

\begin{figure}[htbp]
\epsfxsize=6cm
\centerline{\epsfbox{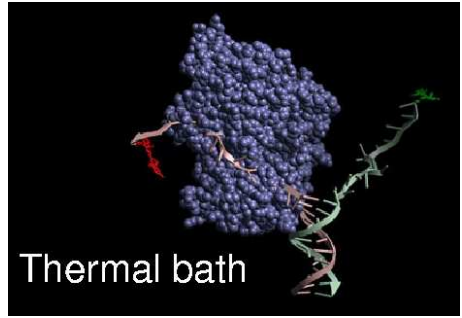}}
\caption{The helicase is an enzyme involved in the unwinding of the DNA
helix that paves the way for the replication process carried out by the
DNA polymerase. Its behavior is thought to be stochastic and intermittent.}
\label{fig1}
\end{figure}

One of the most crucial aspects of many biomolecules (such as RNA
molecules and proteins) is their capability to function as molecular
machines or {\em Maxwell demons} that perform specialized molecular
tasks under nonequilibrium conditions~\cite{Maxwell03}. Often the
innermost workings of such machines is poorly understood, however one
common aspect is their non-deterministic behavior (contrary to the
workings of macroscopic machines). The surrounding water is the thermal
bath by allowing biomolecules to exchange energy with the molecules of
the solvent through the breakage of weak molecular bonds. The amount of
energies typically exchanged during the excursions of the molecular
machine correspond to those delivered in collisions between the
molecules of the solvent and the atoms in the biomolecule that trigger
the relevant conformational changes. Considering that each molecule of
the solvent carries {\em circa} $1k_BT$ ($k_B$ being the Boltzmann
constant and $T$ the temperature of the bath) then the energies
exchanged amount to a few times $k_BT$. This number is roughly equal to
the number of weak bonds that must be broken to trigger the
conformational change. For example, during the replication of DNA, the
replication fork advances one base pair (about 1/3 of a nanometer) every
time the DNA polymerase (a molecular machine) adds one nucleotide to the
newly synthesized DNA strands. The forces that keep the polymerase
moving during this nonequilibrium process are generated from ATP
consumption during the hydrolysis cycle. Often molecular machines do not
act alone but rather a multiplex of several proteins are involved in the
most basic tasks carried out inside the cell. For example, in the
aforementioned case of DNA replication, there is a forerunner of the DNA
polymerase, known as the helicase. The taks of this enzyme is the
progressive unwinding of the double helix as the polymerase advances
(see Fig~\ref{fig1}). Sustained by ATP consumption the helicase exerts
mechanical work upon the DNA, a by-product of the mechanical torque
exerted on the helix and the angle of unwinding required for the
exposure to the polymerase of the successively unwounded base pairs.


Despite the enormous complexity of the whole replication process it is
however interesting to ask how each of these individual motors {\em
work} (e.g. in the case of the helicase) and what is the amount of
energy consumed at periodic time intervals. Surely enough this
quantity will strongly fluctuate as the behavior of these machines is
stochastic and ATP consumption is not deterministic. Although energy
consumption is a rather tricky quantity, mechanical work turns out to
be experimentally measurable as single molecule techniques allow us to
measure forces (or torques) and distances (or angles). Mechanical work
is also a stochastic quantity so we may ask what is the distribution
of the work done by the helicase and measured along many time
intervals of a given duration. For macroscopic machines, if
stochasticity was experimentally observable, we might expect a work
distribution dominated by an extremely narrow Gaussian component as
predicted the law of large numbers. However, for small machines the
distribution might be strikingly different as their inner workings is
a by-product of progressive evolution after millions of years. Quite
probably, the distribution will be strongly non-Gaussian and
intermittent~\cite{Ritort04}, an economy saving strategy for
information transfer~\cite{Loewe99}. This means that most of the time
the helicase does nothing while jiggling at a fast frequency around
its local equilibrium position. However, from time to time and at a much
lower frequency, the helicase hydrolyzes one ATP molecule and makes a
conformation change that triggers the unwinding of an additional base
pair.

The discipline that investigates the thermal behavior of small systems
under various nonequilibrium conditions goes under the name of
nonequilibrium thermodynamics of small systems~\cite{Ritort03,PT}. It addresses the question about the
statistical description of energy exchange processes in small
nonequilibrium systems embedded in thermal environments where the
relevant exchanged energies are few times ($N$) $k_BT$ so relative
deviations (of order $1/\sqrt{N}$) are not negligible over timescales
relevant to biomolecular processes.  The plan of the paper is as
follows. In Sec.~\ref{small} I describe few concepts that are central
in a thermodynamic description of nonequilibrium small systems. In
Sec.~\ref{fts} I briefly discuss the usefulness of
fluctuation-theorems to describe energy exchanged fluctuations in
nonequilibrium processes. Sec.~\ref{sm} describes single molecule
experiments as a promising route to investigate such
fluctuations. Finally I show some recent results regarding work
fluctuations in the mechanical unfolding of RNA molecules
(Sec.~\ref{exps}).

\section{Small systems: heat, work and fluctuations}
\label{small}
A central notion in thermodynamics of small systems is the concept of
{\em control parameter}. This is akin to the concept
of external variable used to define ensembles in statistical
mechanics. The main difference between a thermodynamic description of
macroscopic and small systems is that, in the former, fluctuations are
not essential to characterize thermodynamic transformations.  When
fluctuations are included it is only by considering small deviations
(typically Gaussian distributed) around the average macroscopic
value. Instead, large non-Gaussian deviations are irrelevant as they
are extremely unlikely. For small systems a description in terms of
average values does not suffice, in particular when describing
nonequilibrium thermal processes where rare and large deviations
often occur. When embedded in a thermal environment every observable
of a small equilibrated system strongly fluctuates. In order to define
an equilibrium state it is then convenient to specify the control
parameter.  This is a non-fluctuating quantity that, once fixed,
determines the fluctuating spectrum of the other variables. At
difference with macroscopic thermodynamics, many different equilibrium
states can exist for small systems, depending on which parameter is
externally controlled.


\begin{figure}[htbp]
\epsfxsize=6cm
\centerline{\epsfbox{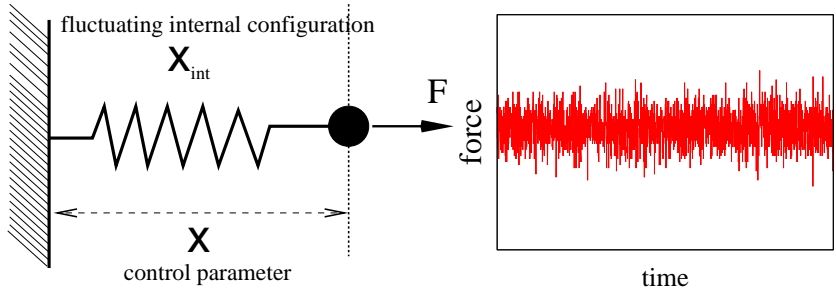}\epsfxsize=5cm\epsfbox{force.time.eps}}
\caption{Schematic picture of a small spring in contact with a bead held at distance $x$ to
the wall. The force exerted on the bead fluctuates with time (right
panel), the spectrum of force-fluctuations being Gaussian for small
deformations of the spring.}
\label{fig2}
\end{figure}
In order to better understand the meaning of the control parameter, let
us think of the following {\em Gedanken} experiment. In Fig.~\ref{fig2}
we show a small bead connected to the extreme of an overdamped spring
whose other extreme is held fixed to a wall. The whole system is
embedded in a thermal bath kept at a given temperature and
pressure. However, at difference with macroscopic systems, we will
assume that the spring is small and made out of few hundreds of atoms
(e.g. this could be a polymer made out of few hundreds of monomers). The
configuration of the system is then specified by an internal set of
variables $\lbrace x_i\rbrace$ specifying the positions of all atoms of
the spring as well as the bead.  In equilibrium, and in absence of any
other interaction with the external world, the extension of the spring
will fluctuate around a reference value that we take equal to zero. If
we want to pull the spring there are different ways we can do that. One
way would be to pull the bead by moving the distance $x(t)$ in a
controlled way. In this case $x$ is the control parameter and the
internal configuration of the spring and the force acting on the bead
will fluctuate (Fig.~\ref{fig2}). For arbitrary deformations (described
by $x$) the average force acting on the bead will satisfy $<F>=f(x)$,
$f$ being a given function with $f'(0)=k$, equal to the stiffness of the
spring. On the other hand, we could pull the spring by controlling the
force (e.g. by applying a external magnetic field to a magnetized
bead). In this case the force would be fixed but the distance $x$ would
fluctuate and satisfy, $F=g(<x>)$ with $g$ another function. In general,
$f\ne g$ so the equilibrium state is different in both protocols
(distance or force controlled). Only for macroscopic systems $f=g$ and
both setups are equivalent.

Let a system be described by an internal configuration
$\lbrace x_i\rbrace$ and a control parameter that we will denote as $x$
(in general there can be a finite number of control parameters). Let
$U(\lbrace x_i\rbrace,x)$ describe the internal energy of the
system. Upon variation of $x$  the energy will change,
\be
dU(\lbrace x_i\rbrace,x)=\sum_i\Bigl(\frac{\partial U}{\partial
x_i}\Bigr)_xdx_i+\Bigl(\frac{\partial U}{\partial x}\Bigr)_{\lbrace x_i\rbrace}dx=dQ+dW
\label{eq1}
\ee
which is the content of the first law of the thermodynamics (i.e. energy
conservation). Now let us consider a process where the spring, initially
in thermal equilibrium at $x=0$, is pulled by changing 
the control parameter according to a perturbation protocol $x(t)$ in a
process that lasts for a time $t_f$ ($x(t_f)=x_f$). If the speed
$\dot{x}$ is much larger than the relaxation frequency of the system $\omega=k/\gamma$
($\gamma$ being the friction coefficient of the bead), then the system will
be driven out-of-equilibrium during the process. The
total work done on the system is given by,
\be
W=\int_0^{x_f} F(\lbrace x_i\rbrace,x)dx
\label{eq2}
\ee
where $F(\lbrace x_i\rbrace,x)$ is the fluctuating force acting upon the
bead,
\be
 F(\lbrace x_i\rbrace,x)=\Bigl(\frac{\partial U}{\partial
 x}\Bigr)_{\lbrace x_i\rbrace}~~~.
\label{eq3}
\ee
If we repeat this nonequilibrium experiment many times always starting from
the same equilibrated state at $x=0$ and following the same protocol
$x(t)$, the system will follow different
trajectories (i.e. the time evolution of $\lbrace x_i\rbrace$ and
therefore the force \eq{eq3} will change from experiment to
experiment. Consequently, the total
work \eq{eq2} will also fluctuate from experiment to experiment. A
quantity that characterizes the nonequilibrium process is the
probability distribution $P(W)$ of work values obtained along
different trajectories. The discussion of some of the mathematical properties of this
distribution is the main subject of concern in this paper, the
quantity that characterizes the small system during the
nonequilibrium process and a fingerprint of its nonequilibrium
behavior.


\begin{figure}[htbp]
\epsfxsize=6cm
\centerline{\epsfbox{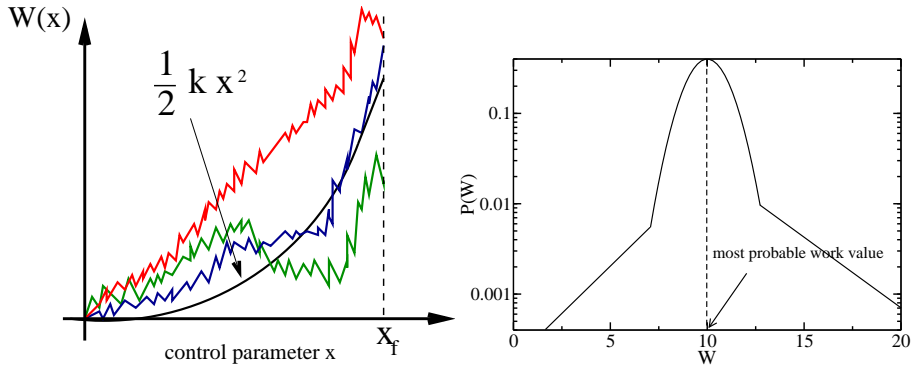}\epsfxsize=6cm\epsfbox{fig1.eps}}
\caption{Fluctuations in the work exerted upon a small spring immersed
in a thermal bath (left panel). The continuous black line is the
average work for small deformations ($k$ is the stifness of the
spring). In general, the probability distribution of the work exerted
upon the system along many repeated experiments (right panel) will
have two sectors characteristic of intermittent behavior: a large
Gaussian component describing small and frequent fluctuations and
exponential tails describing large and rare deviations. Only for
linear systems (i.e. for small deformations) fully Gaussian behavior
is recovered~\protect\cite{Ritort04}.}
\label{fig3}
\end{figure}

\section{Free-energy recovery from nonequilibrium experiments}
\label{fts}
In a nonequilibrium process the second law of thermodynamics~\cite{Fermi56}
establishes that the average work over all trajectories $<W>=\int
WP(W)dW$ is larger than the reversible work (equal to the free-energy
difference $\Delta G$ between the equilibrium states defined at $x=x_f$
and $x=0$). If we define $W_{\rm dis}=W-\Delta G$ as the dissipated work
along a given trajectory, the second law can be
written as,
\be
<W>\ge \Delta G~~~~\rightarrow~~~~<W_{\rm dis}>\ge 0~~~~.
\label{eq4}
\ee
The equality occurs only when the perturbation process is carried out
infinitely slowly in a quasi-static process where $\dot{x}\to 0$. In
such a process the system is given enough time to relax to equilibrium
at each value of the control parameter, therefore $W=\Delta G$ and
$P(W)=\delta(W-\Delta G)$ or $P(W_{\rm dis})=\delta(W_{\rm dis})$ (this
is true for stochastic but not for deterministic
dynamics). Nonequilibrium processes are characterized by hysteresis
phenomena, the average work performed upon the system differs between a
given process and its time-reversed one. Fluctuation theorems assert
relations between the entropy production along a given process (usually
termed as forward) and their reversed one~\cite{Ritort03,EvaSea02}. In
the aforementioned example of the spring, let $x_F(t)$ stand for the
forward protocol that pulls the spring from $x=0$ to $x_F(t_f)=x_f$. The
time reversed protocol is then defined by $x_R(t)=x_F(t_f-t)$. Under the
assumptions that the system is microscopically reversible (detailed
balance) and that the system starts at equilibrium at $x=0$ in the forward
process and at $x=x_f$ in the reverse process, the following result has
been derived by Crooks~\cite{Crooks98}
\be
\frac{P_F(W)}{P_R(-W)}=\exp\Bigl( \frac{W-\Delta G}{k_BT}\Bigr)~~~~,
\label{eq5}
\ee
where $P_F(W),P_R(-W)$ are the work distributions along the forward
and reverse processes respectively (the minus sign in the argument of the reverse
work distribution arises from the corresponding reverse sign of $dx$ in \eq{eq2}). 
Eq.\eq{eq5} has the form of a fluctuation theorem (FT) and quantifies the
amount of hysteresis for arbitrary nonequilibrium protocols. The
quasi-static process is a particular case of \eq{eq5}
where $W=\Delta G$ and there is no hysteresis between the forward and the
reverse paths.

A straightforward consequence of the Crooks FT is the Jarzynski
equality (JE)~\cite{Jarzynski97}. By rewriting \eq{eq5} and integrating out the distribution
$P_R(-W)$ over $W$ it is possible to derive the following
expression 
\be
\langle \exp\Bigl(-\frac{W}{k_BT} \Bigr)\rangle_F=\exp\Bigl( -\frac{\Delta
G}{k_BT} \Bigr)~~~{\rm or}~~~\langle \exp\Bigl(-\frac{W_{\rm dis}}{k_BT} \Bigr)\rangle_F=1~~~~,
\label{eq6}
\ee
where the average $\langle...\rangle_F$ is taken over all possible
work values along the forward process. A consequence of JE is the
second law $\langle W_{\rm dis}\rangle_F\ge 0$ that can be derived by
applying Jensen's inequality ($\langle\exp(x)\rangle\ge \exp(\langle
x\rangle)$). The content of the JE is that, albeit the average
dissipated work is positive, tails in the work distribution that
extend to the region $W_{\rm dis}<0$ must exist for the equality to be
satisfied. Trajectories contributing to these tails are often called
{\em transient violations of the second law} because they violate the
inequality \eq{eq4} for a single trajectory. It has to be stressed,
however, that no violation of the second law occurs as the content of
the inequality only concerns the average value of the work rather the
value of the work of individual trajectories.  The validity and
consistency of the Crooks FT and the JE have been recently put under
scrutiny~\cite{LipDumSmiTinBus02,Ciliberto05,CohenChris04}. Recently, the experimental validity of such
theorem has been tested in RNA pulling
experiments~\cite{Ritort04b} in the far from equilibrium regime and represents an important step in
our understanding of fluctuations in small systems.

Additional interest in the Crooks FT and the JE stems from the fact that
these results can be used to recover equilibrium free-energy differences
from nonequilibrium experiments. This has applications in numerical
simulations of molecular reactions which often cannot be investigated
using equilibrium methods~\cite{Schul04}, or single molecule experiments
where free-energy measurements cannot be carried out
reversibly~\cite{LipDumSmiTinBus02,Ritort03}. In fact, rewriting \eq{eq6} as follows
\be
\Delta G=-k_BT\log\Bigl(\langle \exp(-\frac{W}{k_BT})\rangle  \Bigr)~~~~,
\label{eq7}
\ee
shows that by exponentially averaging the nonequilibrium work it is
possible to recover the value of the reversible work (equal to the
free-energy difference). As always there is no free lunch, and the
main disadvantage of \eq{eq7} lies on the fact that the average
$\langle...\rangle$ must be taken over an infinite number of
nonequilibrium trajectories. The number of available trajectories is
always finite, therefore the risk exists that some of the trajectories
which mostly contribute to the exponential average are not picked
out. Indeed, this is precisely what happens, as the most improbable
trajectories that populate the negative tail of the work distribution
are the ones that mostly contribute to \eq{eq7}. How many
nonequilibrium experiments are needed in order to recover the
free-energy within a given accuracy is one of the most useful
questions one would like to answer. It can be shown that the
exponential average in \eq{eq7} is a biased
quantity~\cite{ZucWoo02,GorRitBus03}, and such number of experiments
increases exponentially fast with the average value of the dissipated
work~\cite{RitBusTin02}. Nevertheless, the precise value of the
prefactor and the factor in the exponential depend in a complicated
way on the left tails of the work distribution.
\begin{figure}[htbp]
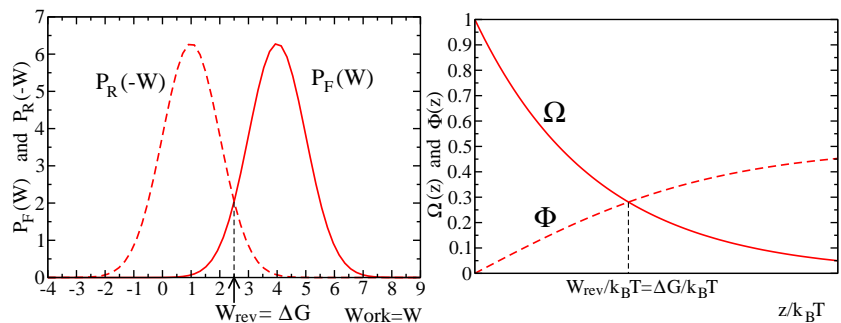

\epsfxsize=5.5cm
\centerline{\epsfbox{cross1.eps}\epsfxsize=5.5cm\epsfbox{cross2.eps}}
\caption{Crossing methods to determine $\Delta G$ from nonequilibrium
work measurements.}
\label{fig4}
\end{figure}


The Crooks FT can also be used for free-energy recovery by applying the
so called crossing methods~\cite{Ritort04b}. Indeed, from
\eq{eq5} we infer that for $W=\Delta G$ both
distributions (forward and reverse) cross each other allowing to extract
the value of $\Delta G$,
\be
P_F(W)=P_R(-W)~~~~~~\rightarrow W=\Delta G
\label{eq8}
\ee
Further improvement of the crossing method uses information from both
distributions along the whole work-axis rather than only local behavior
around $W=\Delta G$. To this end we consider the two functions
\be
\Omega(z)=\frac{N_R(-W>z)}{N_F(W>z)}~~~~;~~~~\Phi(x)=\langle\exp\Bigl(-\frac{(W-z)}{k_B T} \Bigr)\rangle_{F, W>z}~~~~~~,
\label{a4}
\ee
where $N_F(W>z),N_R(-W>z)$ indicate the fraction of trajectories with
work values larger than $z$ along the unfolding and refolding paths
respectively. The average $\langle ...\rangle_{F, W>z}$ is restricted
over the set of trajectories along the forward process with work
larger than $z$. These functions satisfy the following properties: a)
$\Omega (z)$ is a monotonically decreasing function starting at $1$
for $z\ll \Delta G$ and decaying to zero for $z\gg \Delta G$; b)
$\Phi(z)$ is a monotonically increasing function starting at $0$ for
$z\ll \Delta G$ which saturates for $z\gg \Delta G$; c) Both functions
cross each other at $z=\Delta G$.  The two methods are exemplified in
Fig.~\ref{fig4} for the case of Gaussian work distributions (this case
corresponds to a bead confined in an optical trap which is dragged
through water~\cite{MazJar99,WanSevMitSeaEva02}).

\section{Single molecule force microscopy}
\label{sm}
As we have seen in the preceding sections, mechanical work plays a
central role in a thermodynamic description of small systems as there
are specific relations that quantify their probability distributions.
From the experimental point of view, force microscopies provide tools to manipulate
individual biomolecules by applying mechanical force at their ends. In
this way it is possible to exert mechanical work upon these molecules
and, by repeated pullings, to determine
experimentally the work probability distribution in a given
nonequilibrium process.

\begin{figure}[htbp]
\epsfxsize=6cm
\centerline{\epsfbox{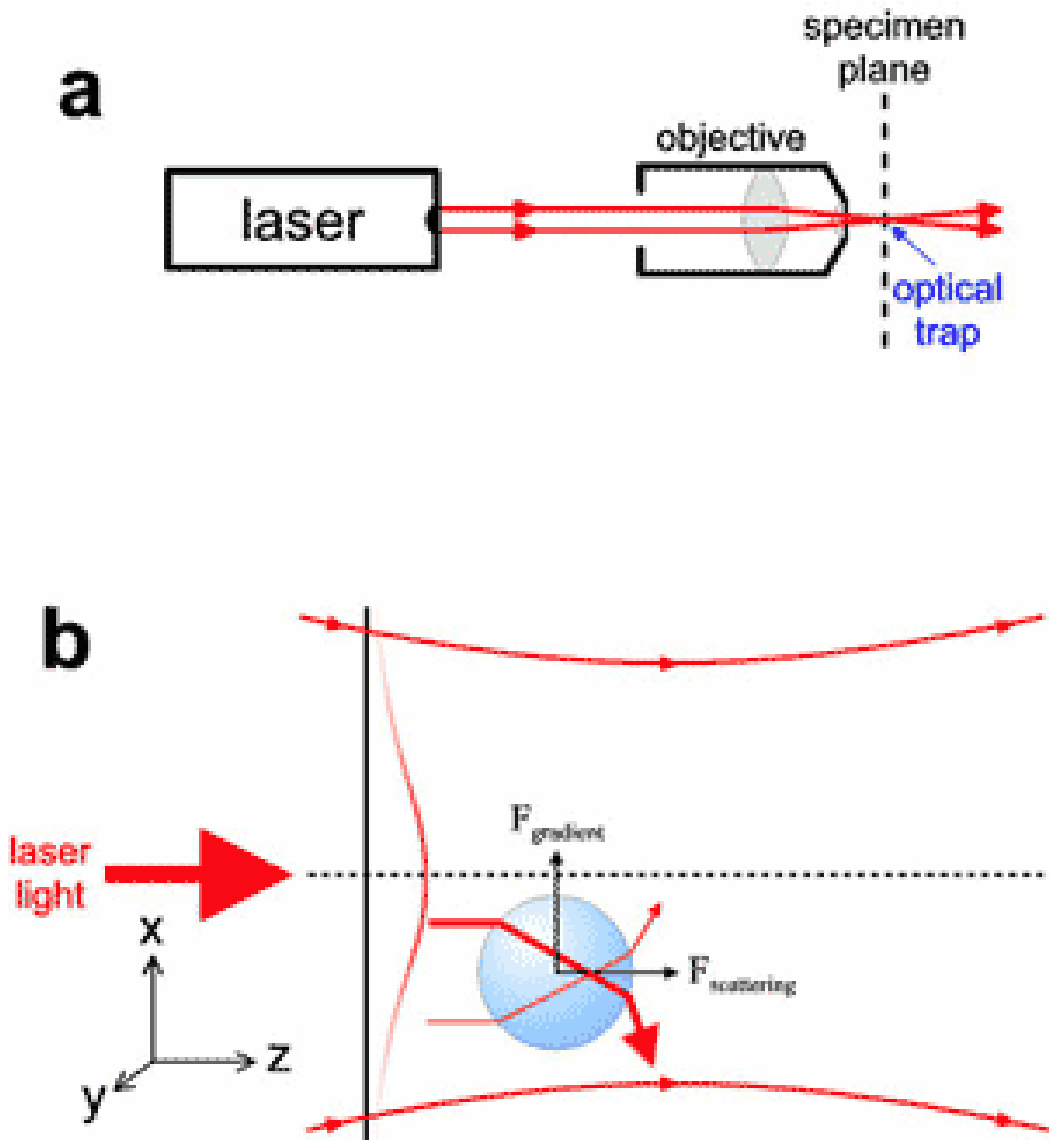}\epsfxsize=8cm\epsfbox{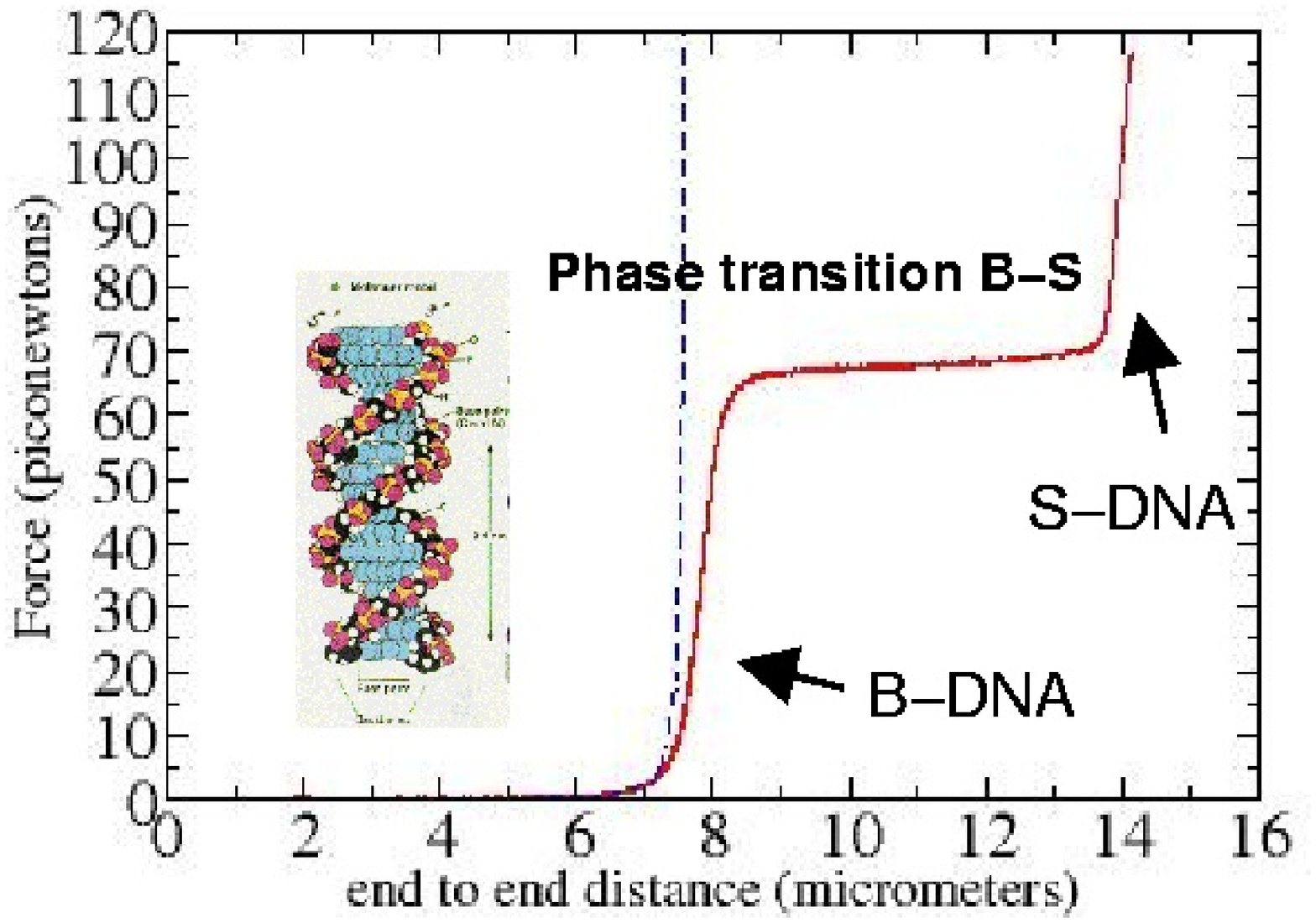}}
\caption{Left panel:Physical principles of the single-beam laser
tweezers. The setup consists of a laser and and an objective (a)
which is focused on a spot. A micron-sized bead is pulled towards
the region of maximum light intensity (b). Right panel: Force-extension
curve (FEC) in a 24kbp fragment of $\lambda$-DNA (torsionally
unconstrained) showing the overstretching transition at 65pN. The
dashed line is the worm-like-chain prediction wich does not include
the elastic rigidity of the backbone.}
\label{fig5}
\end{figure}

\begin{figure}[htbp]
\centerline{\epsfxsize=8cm\epsfbox{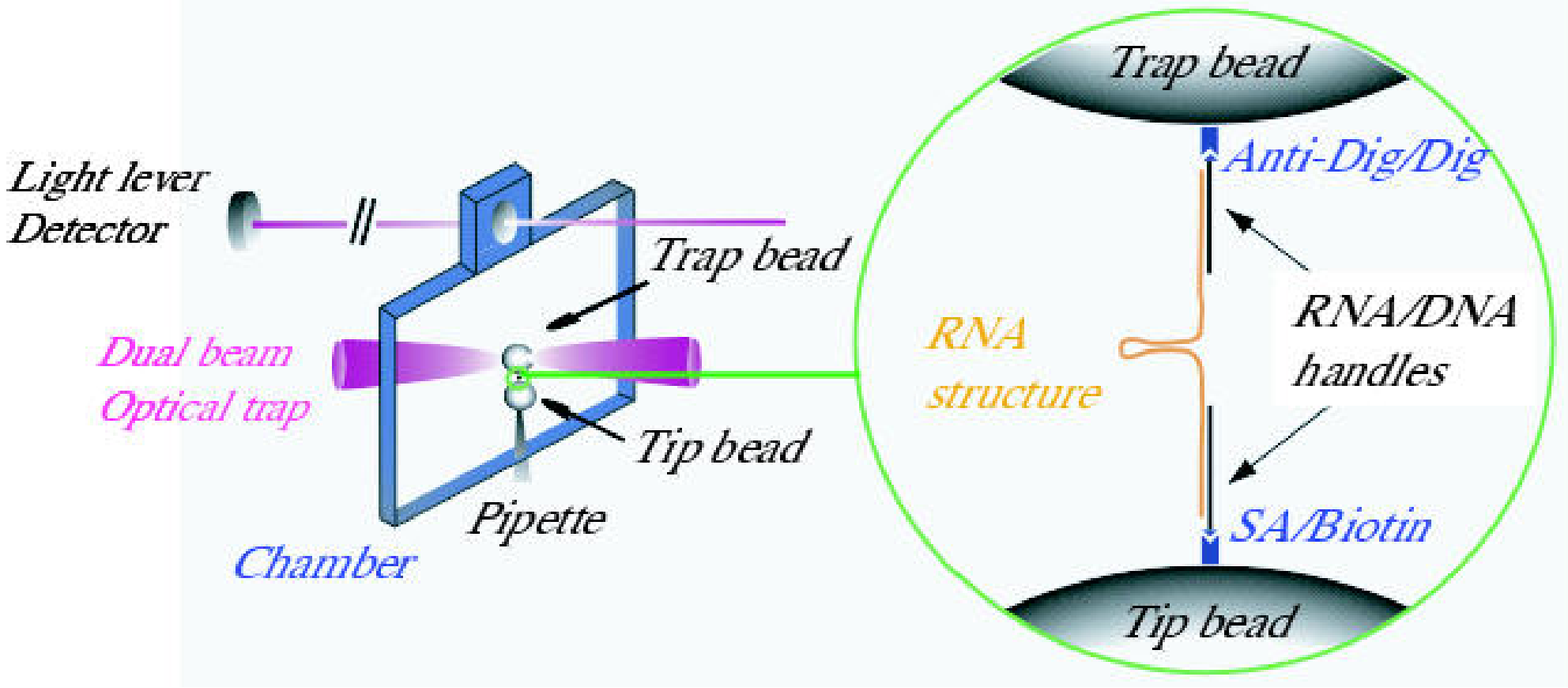}}
\centerline{\epsfxsize=6cm\epsfbox{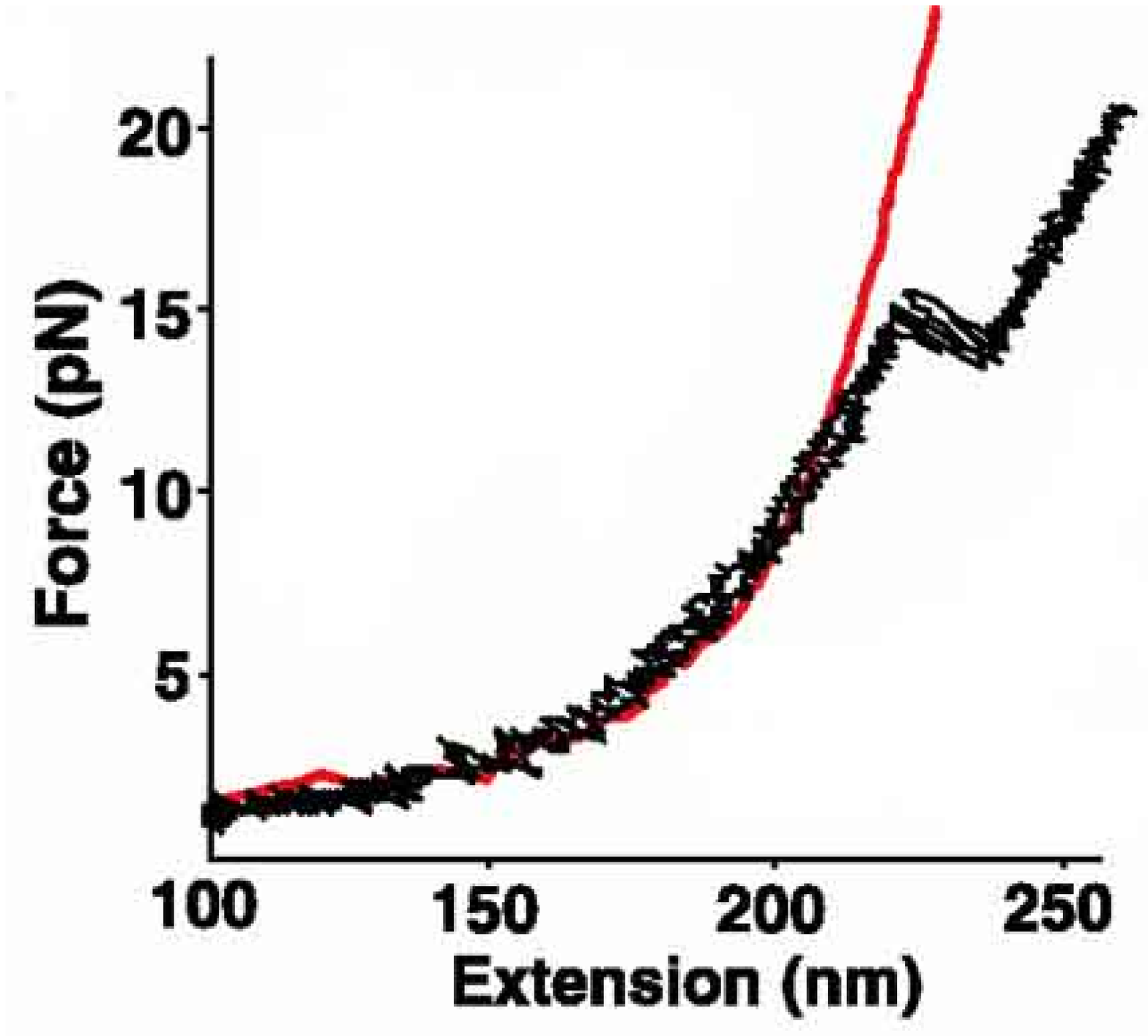}\epsfxsize=6cm\epsfbox{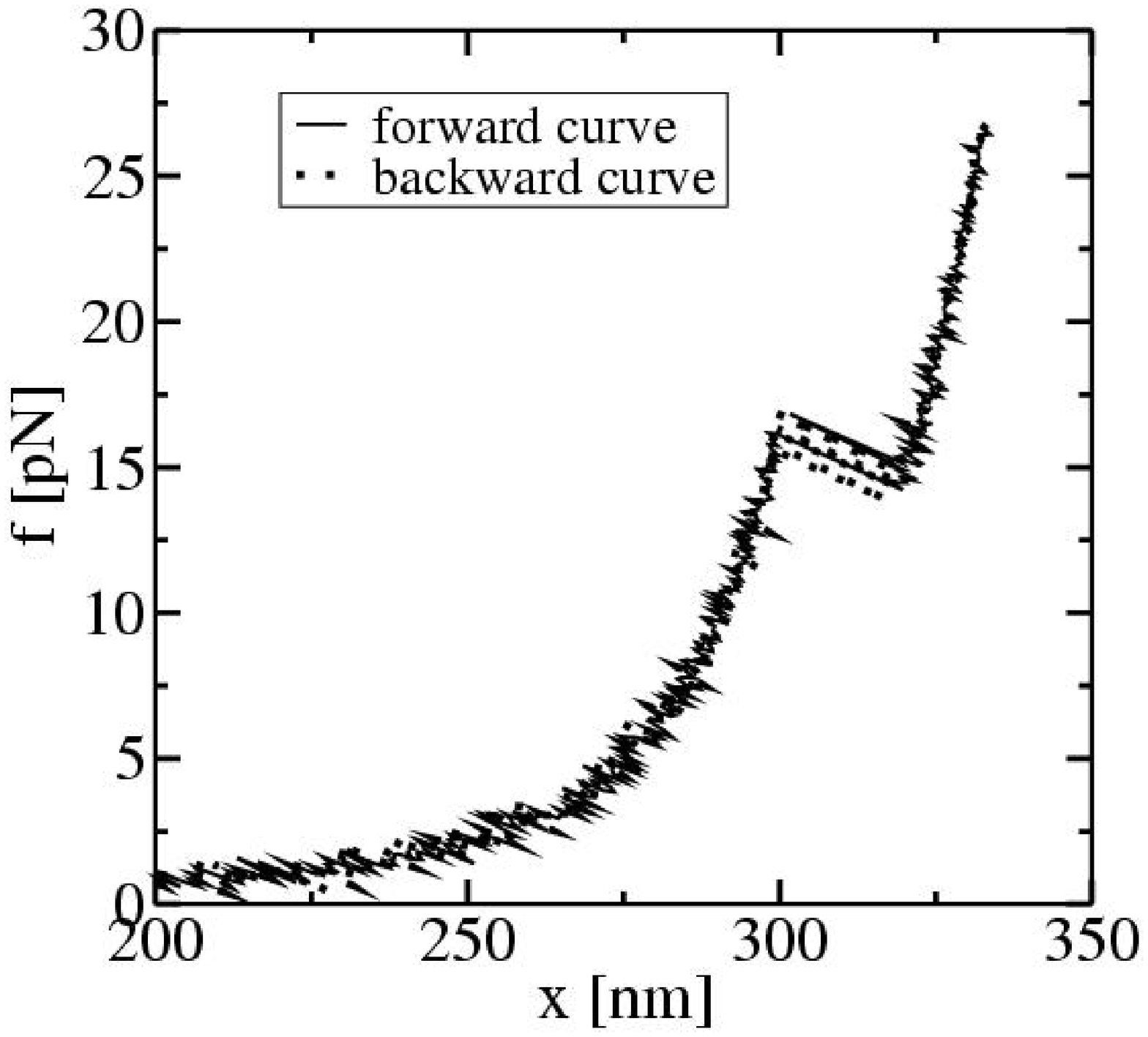}}
\caption{Upper panel: Experimental setup for dual laser tweezers in RNA
pulling experiments. Lower panel: FEC for a small RNA hairpin showing the
rip in the force indicating the unfolding of the molecule. Experiments
have been done in \protect\cite{LipOnoSmiTinBus01} (left panel) and later compared with theoretical models~\protect\cite{Maria04}(right panel).}
\label{fig6}
\end{figure}

There are several kinds of force microscopies, the most well
known are atomic force microscopy, magnetic and optical tweezers. The
latter are particularly suitable as the range of forces they can
exert are in the range 1-100pN relevant to many weak interactions
participating in biomolecular processes. Laser tweezers (see
Fig.~\ref{fig5}) use the principle of conservation of light momentum
to exert forces on small micron-sized polystyrene beads due to light
deflection of the beam as it changes medium between water and the
bead~\cite{SmiCuiBus03}. In this way a bead is trapped into the focus
of the laser, the configuration of minimal energy. When the bead
deviates from the focus a restoring force acts upon the bead, the
principle being the same by which a dielectric substance inside a
capacitor is drawn inwards by the action of the electric field. To a
very good approximation the trap potential is harmonic, therefore the
restoring force acting on the bead is linear with the deviation of the
bead from the center of the trap. Calibration of the optical trap
allows to determine the force acting on the bead by reading the
deviation of the bead from the center of the trap, inasmuch as the
position of the needle in a manometer indicates the value of pressure
of a fluid or a gas. A typical value of the trap stiffness is
0.1pN/nm. In general, it is more convenient to use dual-beam optical
tweezers which do not need continuous calibration as the force can be
determined by the total amount of light collected by two
photodetectors sitting at opposite sides of the beams (see
Fig.~\ref{fig6}).  Experiments use micron-sized glass chambers filled
with water and two beads.  Molecules are chemically labeled at their
ends and polystyrene beads are chemically coated to stick to the ends
of the labeled molecule. In this way a tether can be made
between the two beads. One bead is held fixed by air suction on the
tip of a glass micro-pipette, the other is trapped in the focus of the
laser and used to measure the force applied to the molecule. A
frame-grabber and a light-lever then measure the extension of the
molecule with a precision down to the nanometer range.

The outcome of these experiments are the so called force-extension
curves (FECs) where the force acting on the molecule is represented as
a function of the end-to-end distance between the two beads. In this
way it has been possible to experimentally check that DNA behaves as
some polymer theories predict \cite{Strick03}. At small forces (below
1pN) the polymer behaves like a Hookean entropic spring as described
by the freely jointed chain model~\cite{SmiFinBus92}. At larger forces
deviations occur and the FEC is well described by the worm-like chain
model. Above 5pN enthalpic contributions due to the finite rigidity of
the sugar-phosphate backbone start to be important. Finally, at 65pN a
force plateau is observed characteristic of a transition between the
B-DNA form and a stretched new form of DNA (termed as
S-DNA)~\cite{CluLebHelLavVioChaCar96,SmiCuiBus96} (see
Fig.~\ref{fig5}). FECs provide insight into the inner-workings of
biomolecules. A case of much interest regards the unfolding of RNA
molecules or proteins under the action of mechanical force. Under
physiological conditions these molecules are in a folded or native,
functionally active, conformation. Upon heating or chemical treatment
they denaturate and degrade into an extended, functionally inactive,
conformation. The thermodynamic stability of the native state is
determined by the free-energy difference $\Delta G$ between the two
conformations.  Upon the action of mechanical force RNA hairpins
denature as revealed by the presence of a rip in the
FEC~\cite{LipOnoSmiTinBus01}, see Fig.~\ref{fig6}. These experiments
allow us to obtain estimates of $\Delta G$ by measuring the mechanical
work exerted upon the molecule across the transition. In addition,
hopping effects between the folded and the unfolded conformations also
yield valuable information about the kinetics of unfolding in the
presence of force~\cite{CocMarMon03}, a process thought to be relevant
during the synthesis of proteins (in the translation-elongation
process) in the ribosome. Because of the short extension of the
unfolded hairpins (few tens of nanometers) as compared to the size of
the beads, the experimental setup in RNA pulling experiments is a bit
more elaborated than when pulling DNA (see Fig.~\ref{fig6}). To
harness the RNA molecule two RNA/DNA hybrid handles (typically a few
hundred nanometers long) are attached to its ends.  These handles act
as transducers of the force and have direct influence on the unfolding
kinetics of the RNA molecule. A proper inclusion of all the elements
in the experimental setup (such as the bead in the trap and the
handles) and the correct identification of the control parameter are
required to extract information about the RNA molecule (e.g. the value
of $\Delta G$ or the kinetic rates)~\cite{Maria04}.

\begin{figure}[htbp]
\centerline{\epsfxsize=6cm\epsfbox{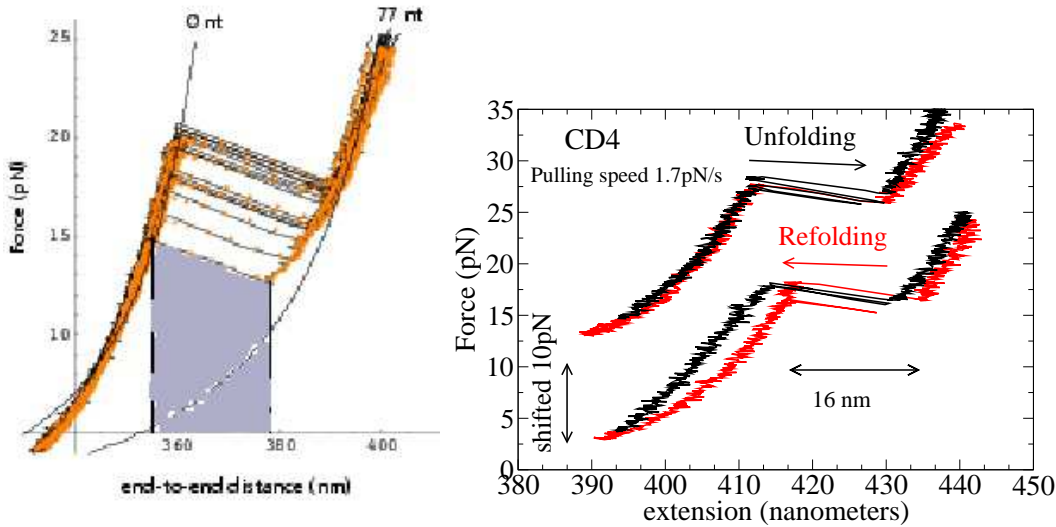}\epsfxsize=8cm\epsfbox{RNA1.eps}}
\caption{Left panel: Pulling curves in the S15 three-way junction
exhibit the strong work fluctuations observed (measured by the gray
area under the FEC). Right panel: Drift effects in the
quasi-reversible unfolding of a short RNA hairpin. For sake of
clarity the second pulling cycle has been shifted 10pN upwards.}
\label{fig7}
\end{figure}


Picking up the threads of our main theme, RNA molecules are specially
suitable to measure work fluctuations. The reason lies in
their modular structure where large RNA molecules are made out of
different motifs or units that unfold sequentially upon pulling \cite{OnoDumLipSmiTinBus03}. The
value of $\Delta G$ for each structural motifs is typically a few tens
of $k_BT$ and each one usually dissipates a few $k_BT$ when unfolded
irreversibly. This quantity is small enough for work fluctuations, as
determined by the value of the force at which the rip occurs, to be
experimentally observable. In Fig.~\ref{fig7} we show work fluctuations as
measured from the area below the FEC ~\footnote{Strictly speaking the
work in RNA pulling experiments is not determined by \eq{eq2} with
$x$ equal to the end-to-end distance but rather by the distance of the
micro-pipette to the center of the trap, yet this difference is too
small as compared to other sources of experimental error to be
significant}.

\section{Predicting unfolding free-energies of RNA motifs from
irreversible measurements of mechanical work.} 
\label{exps}
\begin{figure}[htbp]
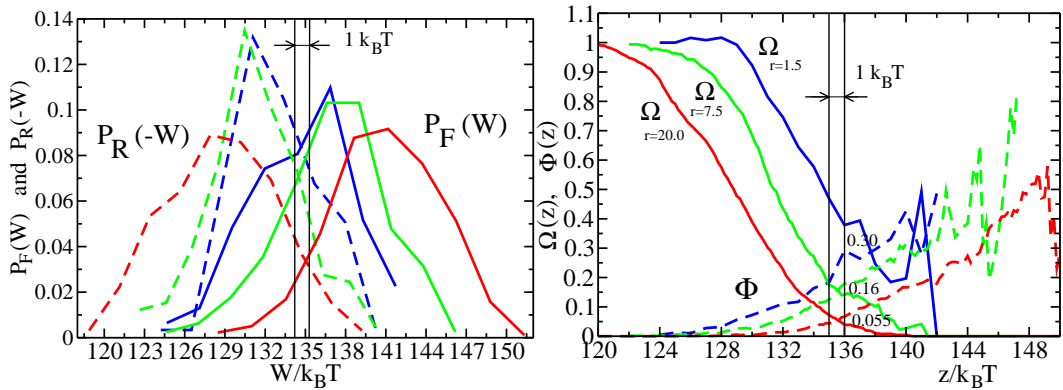

\centerline{\epsfxsize=7cm\epsfbox{pw.new2.eps}\epsfxsize=7cm\epsfbox{tftnew2.eps}}
\caption{Crossing methods applied to unfolding and refolding curves to
a small hairpin pulled at different loading rates
(1.5(blue),7.5(green),20.0(red)pN/s). All curves cross at a value
around $W\sim 135k_BT$ which allows us to extract the value of $\Delta
G$. Results obtained from~\protect\cite{Ritort04b}. At the lowest loading rate
(blue) the quality of the data gets worse due to drift effects.}
\label{fig9}
\end{figure}

As we already said, pulling experiments allow us to extract information
about the unfolding chemical reaction both of thermodynamic character
(the value of $\Delta G$) and kinetic (the reaction rate). Here we want
to discuss more about how to extract the value of $\Delta G$ in RNA
molecules. Traditionally, $\Delta G$ is extracted from calorimetry
experiments by integrating the specific heat as a function of the
temperature across the melting transition. However, in contrast to
proteins, some RNA molecules melt at temperatures above the boiling
point of water, precluding the use of calorimetry measurements. It is
therefore convenient to find new routes to extract the free-energy of
the folded state for such molecules. As we have said, the measure of the
reversible work across the transition in pulling experiments would be a
direct measurement of $\Delta G$. Unfortunately, in most interesting
cases (e.g. RNA molecules with tertiary interactions induced in presence
of $Mg^{2+}$ ions), the unfolding reaction is so slow that it cannot be
carried out reversibly at the available lowest pulling speeds (largely
limited by the presence of strong drift effects in the laser tweezers
machine, see Fig.~\ref{fig7}).  Therefore, other strategies must be
envisaged. Of great utility is the use of the Crooks FT discussed in
Sec.~\ref{fts}. In this case, one can use data from nonequilibrium pulls
to infer the value of $\Delta G$ by looking at the value of the work
where the unfolding and refolding distributions cross each other. In
Fig.~\ref{fig9} we show some experimental data obtained in
\cite{Ritort04b} for a small RNA hairpin in the absence of magnesium
showing that the crossing between both distributions does not depend on
the pulling speed as predicted. The value obtained in this way is also
in agreement with estimates obtained by the Mfold program for the
free-energy of the secondary structure of such motif and for the same
buffer conditions.

\section{Conclusions}
The use of single molecule techniques allows us to investigate the
nonequilibrium behavior of biomolecules \cite{Ritort03}. Such study reveals the presence
of strong thermal fluctuations due to the smallness of the typical
energies associated to the physical interactions in biomolecules (of the
order of few tens of $k_BT$). These energies are small enough for large
deviations respect to the average value be experimentally observable and
important in the timescales relevant to many biomolecular processes \cite{PT}.
Weak molecular bonds (Van der Waals non-specific binding, hydrogen bonds
or hydrophobic interactions) are the leading interactions responsible of
many such processes. They are behind molecular recognition and drive the
transfer of information between biomolecules. It is not a casualty that
weak interactions, which induce strong energy fluctuations, dominate the
inner workings of life processes at the molecular level. It is quite
likely that the large and intermittent fluctuations characteristic of
biomolecules play an important role in the way molecular evolution has
reached such exquisite degree of complexity. This facilitates that large
groups of weakly interacting biomolecules cooperate and carry out very
specialized tasks.

Establishing the nature of work and heat fluctuations in biomolecules
seems therefore relevant to understand the principles underlying the
organization of biological matter at the nanoscale. Statistical
physics provides concepts and tools to address such questions and
fluctuation theorems appear as a good playground to elucidate many
aspects concerning thermal fluctuations in nonequilibrium small
systems. Here we have reviewed a few ideas and experiments in this
exciting field of research which combines knowledge coming from
different areas of expertise, ranging from physics to chemistry and
biology. Sure enough we will see exciting scientific developments in
the future that will help us to better understand the nonequilibrium
thermal behavior of small systems.

{\bf Acknowledgments.} I warmly thank Carlos Bustamante, Delphine
Collin, Chris Jarzynski, Jan Liphardt, Maria Ma\~nosas, Steve Smith and
Nacho Tinoco for a very gratifying collaboration on these subjects
over the past few years. I am grateful to Carmen Miguel for a careful
reading of the manuscript. I also thank the Spanish research council
(BFM2001-3525), the Catalan governement (Distinci\'o de la
Generalitat), The European Union (STIPCO network) and the European
Science Foundation (SPHINX) for various financial support.


\begin{thebibliography}{99}
\bibitem{Houches02} {\it Physics of Biomolecules and Cells}, Les Houches,Session LXXV, EDP Sciences: Springer Verlag (2002)

\bibitem{Maxwell03} {\it Maxwell's Demon 2:Entropy, Classical and Quantum information, Computing}, Edited by H. S. Leff and A. F. Rex, Insitute of Physics Publishing, Bristol (2003)


\bibitem{Loewe99} W. R. Loewenstein, {\it The touchstone of life}, Oxford University Press (1999)


\bibitem{Ritort03} F. Ritort, {\it Seminair\'e Poincar\'e} {\bf 2}, 193
(2003); Available at http://www.ffn.ub.es/ritort/publications.html and arXiV:{\bf cond-mat/0401311}

\bibitem{PT} C. Bustamante, J. Liphardt and F. Ritort, The
Nonequilibrium Thermodynamics of Small Systems, {\it Physics Today} {\bf 58},
43 (2005).

\bibitem{Ritort04} F. Ritort, {\it J. Stat. Mech: Theor. Exp.} P10016 (2004).

\bibitem{Fermi56} E. Fermi, {\it Thermodynamics}, Dover Publications (1956)

\bibitem{EvaSea02} D. Evans and D. Searles, {\it Adv. Phys.} {\bf 51}, 1529 (2002).

\bibitem{Crooks98} G. E. Crooks, {\it J. Stat. Phys.} {\bf 90}, 1481 (1998); {\it Phys. Rev. E} {\bf 61}, 2361 (2000)

\bibitem{Jarzynski97} C. Jarzynski, {\it Phys. Rev. Lett.} {\bf 78}, 2690 (1997); C. Jarzynski, in {\em Dynamics of Dissipation}, P. Garbaczewski, R. Olkiewicz, Eds., (Springer, Berlin 2002).


\bibitem{LipDumSmiTinBus02} J. Liphardt, S. Dumont, S.B. Smith, I. Tinoco Jr. and C. Bustamante,  {\it Science} {\bf 296}, 1832 (2002).

\bibitem{Ciliberto05} F. Douarche, S. Ciliberto, A. Petrosyan and
I. Rabbiosi, Europhys. Lett. {\bf 70}, 593 (2005).

\bibitem{CohenChris04} E. G. D. Cohen and D. Mauzerall, {\it
J. Stat. Mech: Theor. Exp}, P07006 (2004); C. Jarzynski, {\it
J. Stat. Mech: Theor. Exp.} P09005 (2004).

\bibitem{Ritort04b}  D. Collin, F. Ritort, C. Jarzynski, S. B. Smith,
I. Tinoco Jr. and C. Bustamante, {\it Nature} {\bf 437}, 231 (2005).

\bibitem{Schul04} S. Park and K. Schulten, J. Chem. Phys. {\bf 120}, 5946 (2004).

\bibitem{ZucWoo02} D.M. Zuckerman and T.B. Woolf, {\it Phys. Rev. Lett.} {\bf
89}, 180602 (2002).

\bibitem{GorRitBus03} J. Gore, F. Ritort and C. Bustamante, {\it Proc. Nat. Acad. Sci. USA} {\bf 100}, 12564 (2003).

\bibitem{RitBusTin02} F. Ritort, C. Bustamante and I. Tinoco Jr., {\it Proc. Nat. Acad. Sci. USA} {\bf 99}, 13544 (2002).

\bibitem{MazJar99} O. Mazonka and C. Jarzynski, {\em Preprint arXiv:cond-mat/9912121}.

\bibitem{WanSevMitSeaEva02} G.M. Wang, E.M. Sevick, E. Mittag, D.J. Searles, and D.J. Evans, {\it Phys. Rev. Lett.} {\bf 89}, 050601 (2002).

\bibitem{SmiCuiBus03} S.B. Smith, Y. Cui and C. Bustamante, {\it Methods. Enzymol.} {\bf 361}, 134 (2003).

\bibitem{Strick03} T. R. Strick, M-N Dessinges, G. Charvin, N. H. Dekker, J-F Allemand, D. Bensimon and V. Croquette, {\it Rep. Prog. Phys.} {\bf 66}, 1 (2003).

\bibitem{SmiFinBus92}  S.B. Smith, L. Finzi and C. Bustamante, {\it Science} {\bf 258}, 1122 (1992).

\bibitem{CluLebHelLavVioChaCar96} P. Cluzel, A. Lebrun, C. Heller, R. Lavery, J.-L. Viovy, D. Chatenay and  F. Caron, {\it Science} {\bf 271}, 792 (1996).

\bibitem{SmiCuiBus96} S.B. Smith, Y. Cui, C. Bustamante, {\it Science} {\bf 271}, 795 (1996).


\bibitem{LipOnoSmiTinBus01} J. Liphardt, B. Onoa, S.B. Smith,
I. Tinoco Jr. and C. Bustamante, {\it Science} {\bf 292}, 733 (2001).

\bibitem{CocMarMon03} S. Cocco, R. Monasson and J. Marko,
{\it Eur. Phys. J. E} {\bf 10}, 153 (2003).


\bibitem{Maria04} M. Manosas and F. Ritort, {\it Biophys. J} {\bf 88},
3224 (2005).

\bibitem{OnoDumLipSmiTinBus03}  B. Onoa, S. Dumont, J. Liphardt, S.B. Smith, I. Tinoco Jr. and C. Bustamante, {\it Science} {\bf 299}, 1892 (2003).


\end{thebibliography}
\end{document}